\documentclass{iau}
\usepackage{graphicx}

\title[Modes of star formation from Herschel]
{Modes of star formation from Herschel}

\author[Testi, Bressert \& Longmore]   
{L. Testi$^{1,2}$,
E. Bressert$^{1,3}$
 \and S. Longmore$^1$}

\affiliation{$^1$ESO, Karl Schwarzschild srt. 2, D-85748 Garching,
Germany, email: {\tt ltesti@eso.org} \\[\affilskip]
$^2$ INAF-Osservatorio Astrofisico di Arcetri, Largo E. Fermi 5, I-50125 Firenze, Italy
\\[\affilskip]
$^3$School of Physics, University of Exeter, Stocker Road, Exeter EX4 4QL, UK}

\pubyear{2013}
\volume{292}  
\pagerange{119--126}
\jname{Title of your IAU Symposium}
\editors{T. Wong \& J. Ott}
\begin{document}

\maketitle

\begin{abstract}
We summarize some of the results obtained from Herschel  surveys of the nearby star forming regions and the Galactic plane. We show that in the nearby star forming regions the starless core spatial surface density distribution is very similar to that of the young stellar objects. This, taken together with the similarity between the core mass function and the initial mass function for stars and the  relationship between the amount of dense gas and star formation rate, suggest that the cloud fragmentation process defines the global outcome of star formation. This ``simple'' view of star formation may not hold on all scales. In particular dynamical interactions are expected to become important at the conditions required to form young massive clusters. We describe the successes of a simple criterion to identify young massive cluster precursors in our Galaxy based on (sub-)millimetre wide area surveys. We further show that in the location of our Galaxy where the best candidate for a precursor of a young massive cluster is found, the ``simple'' scaling relationship between dense gas and star formation rate appear to break down. We suggest that in regions where the conditions approach those of the central molecular zone of our Galaxy it may be necessary to revise the scaling laws for star formation.
\end{abstract}

\firstsection 
\section{Introduction}

The Herschel Space Observatory multiband wide area surveys of our own Galaxy in the far infrared 
and submillimeter, when 
combined with surveys at other wavelengths allow us, for the first time, to obtain a detailed view of the
relationship between the physical structure of gas and dust in molecular clouds and cores and the young
stellar populations that are produced in these clouds. In this contribution we focus on three results using
data from the Herschel Gould Belt Survey and the HIGAL survey. Detailed descriptions of these surveys 
are given elsewhere in this volume by Di Francesco and Molinari \cite[see also Molinari et al.~(2010) and Andr\'e et al.~(2010)]{Molinari:2010,Andre:2010}.

We will focus on the origin of the observed ranges of young stellar objects (YSO) surface densities in the nearby star forming regions, on the possible identification of precursors of the most massive clusters in our own Galaxy and the special conditions of star formation in the Central Molecular Zone (CMZ) of our Galaxy as compared to the Galactic disk, and in particular the star forming regions in the solar neighborhood.

\section{The starless cores surface density distribution}

Following the extensive Spitzer surveys in nearby star forming regions, it has become clear that young stellar objects are found in a continuous range of stellar densities rather than showing a clear bimodal distribution as postulated by the paradigm of clustered vs. distributed star formation. In particular \cite[Bressert et al.~(2010)]{Bressert:2012} showed that the distribution of surface YSO densities in the regions surveyed by Spitzer is scale-free and the definition of the percentage of stars forming in clusters is arbitrary if based on specific thresholds on the local surface density. 

Thanks to the wide area surveyed and the large samples of starless cores detected, the Herschel surveys of nearby star forming regions allow us to investigate whether this continuous distribution of surface densities are an imprint of the molecular cloud fragmentation. In \cite[Bressert et al.~(2012b)]{Bressert:2012b} we combined the 
Herschel data in the Perseus West and Serpens clouds with infrared data from the Wide-Field Infrared Survey Explorer \cite[(WISE, Wright et al. 2010)]{Wright:2010} to select samples of starless
cores and compare their spatial distribution with that of the YSOs. We employed WISE color-color diagrams to identify YSOs and remove extragalactic contamination \cite[(following the method of Koenig et al.~2012)]{Koenig:2012}, this method was cross checked, in the overlap region, with the YSOs classification in the Spitzer c2d survey \cite[(Evans et al.~2009)]{Evans:2009}. 

\begin{figure}[t]
\begin{center}
\includegraphics[height=4.3cm]{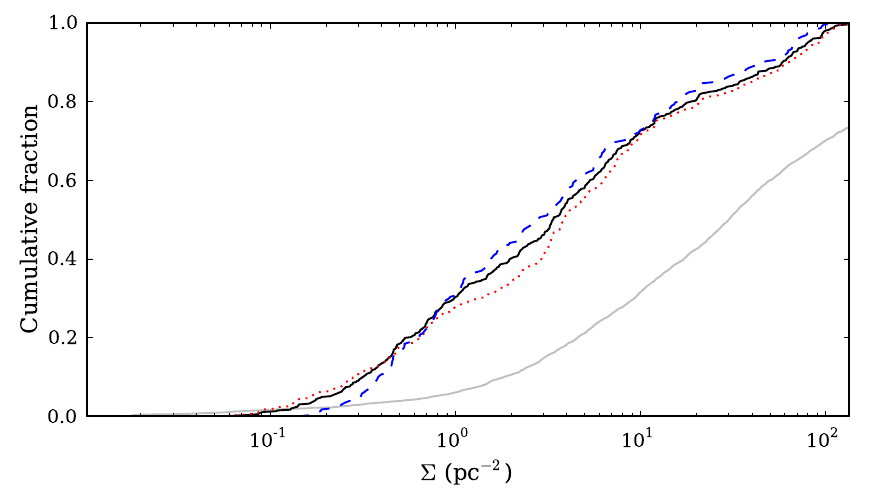} 
 \caption{Cumulative distribution of surface densities of cores and YSOs in Perseus and Serpens combined. Red, dotted line is for Herschel starless cores, blue, dashed line for the WISE YSOs. Solid black line is for the two samples combined. The thin grey line is the distribution for YSOs in nearby star forming regions \cite[(Bressert et al.~2010)]{Bressert:2010}, this reaches higher surface densities but has a shape consistent with the other distributions. Figure adapted from \cite[Bressert et al.~(2012b)]{Bressert:2012b}.}
   \label{fb12b}
\end{center}
\end{figure}

The spatial distribution of cores and young stellar object are being compared using two separate methodologies. We first compare the cumulative distributions of surface densities of the neighbours of each YSO or core and we show that, as found by \cite[Bressert et al.~(2010)]{Bressert:2012} for YSOs, the core distribution is scale free and matches closely that of the formed stars (see Fig.~\ref{fb12b}). In addition, we employed a minimum spanning tree method to identify groups of YSOs or cores, following the method of \cite[Gutermuth et al.~(2009)]{Gutermuth:2009}. As discussed above, given the lack of preferential clustering scales, the identification of groups is based on an arbitrary criterion, but if we use the same criterion for YSOs and cores, we find  that the groups of the two populations tend to overlap spatially and the average densities that we derive are consistent within the uncertainties (albeit affected by very large errors).


These results strongly favors the view in which the density distribution of YSOs in these regions are the result of the imprint of the cloud fragmentation process, rather than dynamical evolution. Taken together with the fact that the core mass function in nearby star forming regions closely resembles the initial mass function for stars \cite[(Testi \& Sargent~1998; Motte, Andr\'e \& Neri~1998; K\"onyves et al.~2010)]{TS98,MAN98,Konyves:2010}, these results strongly support the view that in these regions the global outcome of the star formation process is mainly determined by cloud fragmentation process rather than by dynamical interactions.

\section{Looking for Young Massive Clusters precursors}
 
Nearby star forming regions do not have the conditions to form very massive and dense stellar clusters, and have very modest capabilities of forming  high mass stars (if at all). Nevertheless, the Galaxy is known to host a number of very massive and dense clusters, which are also prominently seen in extragalactic starbursts. These clusters are normally called Young Massive Clusters (YMC) or, in the most extreme cases, Super Star Clusters and have total stellar masses and stellar densities that in some cases approaching those of globular clusters \cite[(see Portegies Zwart et al.~2010, for a recent review)]{PZ2010}. The conditions to form these systems are expected to be very different than those found in nearby star forming regions and dynamical interactions during formation may be the dominant mechanism \cite[(Smith et al.~2009)]{Smith:2009}. Given the importance of these systems for extragalactic studies, constraining the physical conditions for their formation is especially important. The precursors of YMCs are expected to be very rare in our galaxy, but wide area surveys of our Galaxy at far infrared and millimetre wavelengths offer a unique possibility to search for these rare objects. 

\begin{figure}[t]
\begin{center}
\includegraphics[height=4.3cm]{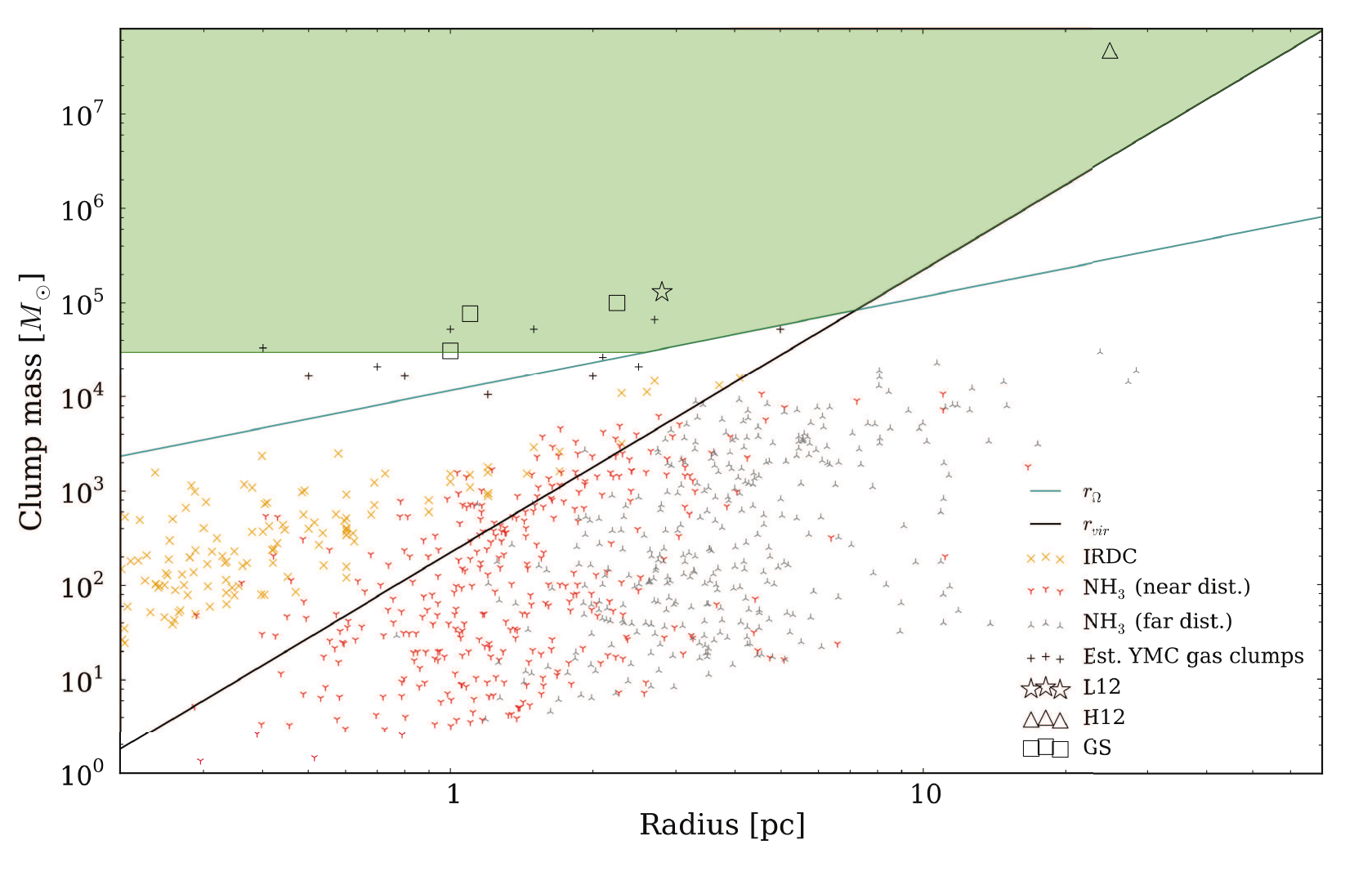} 
 \caption{Criteria to identify precursors of YMCs on the radius-mass plane. The values for $r_\Omega$ and $r_{vir}$ are shown as a thin blue line and a thick black line respectively (see text for definitions). The open symbols represent candidates YMC progenitors: triangle is a YMC candidate progenitor in the Antennae Galaxies \cite[(Herrera et al.~2012)]{Herrera:2012}; the star is G~$0.25+0.02$ near the Galactic Centre \cite[(Longmore et al.~2012a)]{Longmore:2012a}; the squares are additional candidates in our Galaxy from \cite[Ginsburg et al.~(2012)]{Ginsburg:2012}. Figure adapted from \cite[Bressert et al.~(2012a)]{Bressert:2012a}}
   \label{fb12a}
\end{center}
\end{figure}

In \cite[Bressert et al.~(2012a)]{Bressert:2012a} we developed and tested a set of guidelines to identify good candidates of YMC precursors. The criteria that we developed are based on the fact that these precursors not only need to contain enough mass to form a YMC, but also need to be very compact to allow for rapid and continuous star formation even under the effect of strong feedback from forming very massive stars. The final criteria to form a bound YMC with M$_{tot}\ge10^4$~M$_\odot$, requires objects to have  more  than $3\times 10^4$~M$_\odot$ (assuming 30\%\ star formation efficiency), and a radius smaller than the minimum between $r_\Omega$, the maximum radius needed to gravitationally bind a fully ionised core, and $r_{vir}$, the maximum radius that would allow a crossing time of $\sim$1~Myr \cite[(the typical observed value for YMCs, Portegies Zwart et al.~2010)]{PZ2010}. 

These criteria successfully confirm the earlier identifications of candidate YMC precursor identified combining ALMA and VLT data in the Antennae Galaxies \cite[(Herrera et al.~2012)]{Herrera:2012} as well as the candidate precursor G~$0.25+0.02$ identified in the HOPS and HIGAL surveys by \cite[Longmore et al.~(2012a)]{Longmore:2012a}. A few more candidates, identified by \cite[Ginsburg et al.~(2012)]{Ginsburg:2012}, need further study. G~$0.25+0.02$ in our Galaxy offer a unique opportunity to study in detail the initial conditions for YMC formation. The initial results seem to suggest that this YMC precursor is a highly turbulent cloud \cite[(Rathborne et al.~2012)]{Rathborne:2012}. While the properties and internal structure of the cloud are still being investigated, the origin of the cloud is unclear. The cloud is located in the CMZ of our own Galaxy and the special conditions near the Galactic Centre may be responsible for its formation. However, it should be noted that YMCs are not only found in the CMZ in our own Galaxy. Further studies of this and other candidates will help clarify their nature.

\section{Star formation and dense gas in the CMZ}

In the previous section we have discussed a unique massive core in the CMZ of our Galaxy, and we speculated that in the CMZ the conditions may be favorable for the assembly of such extreme objects. In the solar neighborhood, several different arguments suggest that the process of star formation is controlled by the amount of dense molecular gas. \cite[Lada et al.~(2012) and Krumholz et al.~(2012)]{Lada:2012,Krumholz:2012} presented the latest incarnation of the views that the star formation rates observed in star forming regions and entire galaxies could be understood in terms of simple scaling relations with the amount of dense gas above a certain column density threshold (Lada's formulation) or the total local volumetric gas density (Krumholz's formulation).

We used data from HOPS and HIGAL Galactic plane surveys as well as some other datasets to show that the relationships between star formation rate and dense gas that appear to hold throughout the Galactic disk do appear to break down in the CMZ \cite[(Longmore et al. 2012b)]{Longmore:2012}. More details on the methodology and results are given by Longmore in this volume, but the basic result is that while $\sim$80\%\ of the dense gas in our own Galaxy is concentrated in the CMZ, there is no similar enhancement in the star formation rate. Similarly, the predictions of the volumetric gas density relations exceed by at least a factor of ten the observed star formation rate. These findings suggest that the prescriptions based on a single parameter (the total gas density or the dense gas column density) break down in the CMZ and a more complicated formulation is needed to reconcile this region with the rest of the Galaxy. The large shear and increased turbulence in the CMZ need to be looked at as the possible causes for the increased stability of cloud cores in that region. These processes may also favor the formation of more massive stable clumps that once they became unstable and collapse may form YMCs.


\begin{thebibliography}{}

\bibitem[Andr\'e \etal\ (2010)]{Andre:2010}
{Andr\'e, Ph., MenÕshchikov, A., Bontemps, S., et al.} 2010, 
\textit{A\&A}, 518, L102 

\bibitem[Bressert \etal\ (2010)]{Bressert:2010}
{Bressert, E., Bastian, N., Gutermuth, R., et al.} 2010,
\textit{MNRAS}, 409, L54 

\bibitem[Bressert \etal\ (2012a)]{Bressert:2012a}
{Bressert, E., Ginsburg, A., Bally, J., et al.} 2012,
\textit{ApJL}, in press 

\bibitem[Bressert \etal\ (2012b)]{Bressert:2012b}
{Bressert, E., Testi, L., Facchini, A., et al.} 2012,
\textit{A\&A}, submitted 

\bibitem[Evans \etal\ (2009)]{Evans:2009}
{Evans, N. J., Dunham, M. M., J¿rgensen, J. K., et al.} 2009, \textit{ApJS}, 181, 321

\bibitem[Ginsburg \etal\ (2012)]{Ginsburg:2012}
{Ginsburg, A., Bressert, E., Bally, J., Battersby, C., et al.} 2012, \textit{ApJ}, in press

\bibitem[Herrera \etal\ (2012)]{Herrera:2012}
{Herrera,C.N., Boulanger,F., Nesvadba,N.P.H., Falgarone,E.} 2012, \textit{A\&A}, 538, L9

\bibitem[Koenig \etal\ (2012)]{Koenig:2012}
{Koenig, X. P., Leisawitz, D. T., Benford, D. J., et al.} 2012, \textit{ApJ}, 744, 130

\bibitem[K\"onyves \etal\ (2010)]{Konyves:2010}
{K\"onyves, V., Andr\'e, Ph., Men'shchikov, A., et al.} 2010,
\textit{A\&A}, 518,  L106 

\bibitem[Krumholz \etal\ (2012)]{Krumholz:2012}
{Krumholz, M.R., Dekel, A., McKee, C.F.} 2012, \textit{ApJ}, 745, 69

\bibitem[Lada \etal\ (2012)]{Lada:2012}
{Lada, C.J., Forbrich, J., Lombardi, M., Alves, J.F.} 2012, \textit{ApJ}, 745, 190

\bibitem[Longmore \etal (2012b)]{Longmore:2012b}
{Longmore,S.N., Bally, J., Testi, L., et al.} 2012, \textit{MNRAS}, arXiv:1208.4256

\bibitem[Longmore \etal (2012a)]{Longmore:2012a}
{Longmore,S.N., Rathborne,J., Bastian,N., et al.} 2012, \textit{ApJ}, 746, 117

\bibitem[Molinari \etal\ (2010)]{Molinari:2010}
{Molinari, S., Swinyard, B., Bally, J., et al.} 2012,
\textit{A\&A}, 518,  L100 

\bibitem[Motte, Andr\'e \&\ Neri (1998)]{MAN98}
{Motte, F., Andr\'e, Ph., \&\ Neri, R.} 1998,
\textit{A\&A}, 336, 150 

\bibitem[Portegies Zwart, McMillan \&\ Gieles]{PZ:2010}
{Portegies Zwart, S.F., McMillan, S.L.W., \&\ Gieles, M.} 2010, \textit{ARA\&A} 48, 431

\bibitem[Rathborne \etal\ (2012)]{Rathborne:2012}
{Rathborne, J.M., Longmore, S.N., Jackson, J.M., et al.} 2012, \textit{ApJ}, submitted

\bibitem[Smith \etal\ (2009)]{Smith:2009}
{Smith, R.J., Longmore, S.N., Bonnell, I.} 2009, \textit{MNRAS}, 400, 1775

\bibitem[Testi \& Sargent (1998)]{TS98}
{Testi, L. \& Sargent, A.I.} 1998, \textit{ApJ}, 508,  L91 

\bibitem[Wright \etal\ (2010)]{Wright:2010}
{Wright, E. L., Eisenhardt, P. R. M., Mainzer, A. K., et al.} 2010, \textit{AJ}, 140, 1868

%
%
%
%
%
%
%
%
%
%
%
%
%
%
%
%

\end{thebibliography}
\end{document}